\documentclass[11pt, a4paper]{article}
\usepackage{graphicx}
\usepackage{amsmath}
\usepackage[utf8]{inputenc}
\usepackage[margin = 2.5 cm]{geometry}
\usepackage[numbers,compress]{natbib}

\title{Neural Computation Runs on Dissipative Structures Under Cyclic Calibration: A Systems Perspective on Sleep}
\author{
  Javier Antonio D\'{i}az Cisternas \\
  \date{}
  \small International Institute for Integrative Sleep Medicine (IIIS) \\
  \small University of Tsukuba \\
  \footnotesize \texttt{diaz.antonio.fn@u.tsukuba.ac.jp}
}

\begin{document}

\maketitle

\begin{abstract}
Biological function operates far from thermodynamic equilibrium, sustained by metastable structures that continuously cycle through synthesis, function execution, and degradation---consuming energy and exporting entropy in the process. Neural computation exemplifies this dissipative organization, relying on low-entropy structures engaged in dynamic processes that define key information transfer and processing parameters such as gain, threshold, and integration time. This perspective proposes that sleep serves as a systemic recalibration phase required to reset crucial signaling parameters that drift during dissipative processing. During wakefulness, neural activity progressively consumes and disrupts the organization of the very substrate from which it emerges, leading to parameter drift and reduced computational precision. Sleep maintains computational accuracy across activity cycles by recalibrating function-structure coupling---that is, by counterbalancing entropic drift. Rather than explaining the experimentally observed \emph{sleep need} through homeostatic regulation of unknown physiological variables, this perspective reframes it as an emergent property of neural computation under thermodynamic constraints.
\end{abstract}

\section{Introduction}

Sleep remains one of the most ubiquitous yet enigmatic phenomena in biology \citep{Greene2004-zp, Krueger2016-pr, Yanagisawa2019-sm, Fraigne2026-am}. Despite decades of research, a comprehensive theory explaining why sleep is essential and why its absence triggers widespread cognitive and physiological dysfunction remains elusive \citep{Banks2007-kp, Mullington2009-ku, Buxton2012-zw, kryger2022principles, Faraut2012-xl}. Numerous hypotheses have been proposed---the most influential centered on synaptic homeostasis \citep{Tononi2003-dd, Tononi2014-sf}, memory consolidation \citep{Walker2004-aa, Klinzing2019-ce}, metabolic waste clearance \citep{Xie2013-kd, Fultz2019-ye}, and energy balance \citep{Walker1980-ez, Siegel2009-jb, DiNuzzo2017-hz}. Yet, while each model is supported by compelling empirical evidence, a unifying framework capable of reconciling these heterogeneous theories remains elusive.

Sleep is phenotypically characterized as a distinct physiological and behavioral state, readily identifiable by a marked reduction in responsiveness to environmental stimuli and by well-defined electrophysiological patterns \citep{kryger2022principles}, particularly those captured by the electroencephalogram (EEG) \citep{Pereira2025-hg}. Importantly, sleep is homeostatically regulated, evidenced by the compensatory rebound observed following sleep deprivation \citep{Deboer2013-lk}. Indeed, this rebound has become a central behavioral criterion for operationalizing sleep across diverse taxa \citep{Anafi2019-fn}. Yet, paradoxically, the underlying physiological variable under homeostatic control remains unknown \citep{Franken2024-at}.

Sleep and sleep-like states are evolutionarily conserved across the animal kingdom, present even in organisms with simple nervous systems—or, in some cases, no centralized brain at all \citep{Nath2017-en, Kanaya2020-ly}. Despite its widespread occurrence, sleep remains puzzling from an evolutionary standpoint: it imposes immediate vulnerability and substantial opportunity costs. Any satisfactory theory must therefore account for sleep’s universality and identify the fundamental physiological requirement that justifies its persistence despite these fitness costs \citep{Joiner2016-ig, Anafi2019-fn}.

This perspective approaches the mystery of sleep by investigating the biophysical constraints that govern the physical substrates of neural computation. Like all biological systems, neural networks operate far from thermodynamic equilibrium as dissipative structures \citep{nicolis1977self, England2022-xq}. Consequently, they face a distinct challenge: the turnover of biological structure simultaneously drives hardware maintenance and information processing, ultimately constraining computational fidelity. Reframing the problem through these foundational thermodynamic constraints provides a fresh access point for understanding both the origin and the physical necessity of sleep.

\section{Dissipative structures carry out Biological Function}

\subsection{Dissipative Structures: Self-Organization Far from Equilibrium}

In an isolated system, localized energy dissipates down gradients until it is homogeneously distributed, macroscopic fluxes cease, and state variables become static (Fig. 1a). This state of thermodynamic equilibrium captures the universe's baseline tendency to relax toward maximum disorder, providing the classical benchmark against which non-equilibrium phenomena are measured \citep{callen1991thermodynamics, Kondepudi2014-lc}. However, the most compelling phenomena in nature---from stellar dynamics to biological organization---emerge in open systems operating far from equilibrium. When sustained energy and matter fluxes push physical substrates past critical thermodynamic thresholds, simple proportional responses collapse; microscopic fluctuations become amplified through feedback loops, unlocking non-linear interactions and spontaneous symmetry breaking (Fig. 1b). In this regime, physical systems exhibit a remarkable capacity for self-organization, spontaneously forming ordered, dynamic configurations known as \emph{dissipative structures} \citep{prigogine1968introduction, nicolis1977self}.

Unlike static equilibrium structures (such as crystals), which are stabilized by passive potential energy minima, dissipative structures are inherently dynamic, self-sustaining configurations---exemplified by natural phenomena such as typhoons, Rayleigh--Bénard convection cells, reaction--diffusion patterns, and atmospheric/oceanic currents \citep{Michaelian2026-uc}. They function as specialized thermodynamic pathways that organize matter to dissipate driving free energy gradients, sustaining localized, low-entropy states through continuous global entropy production \citep{Kondepudi2020-av}.

\subsection{Far from equilibrium: The thermodynamic asymmetry of biological function turnover}

It is now widely accepted that living systems are composed of, and actively maintained by, dissipative structures \citep{Prigogine_Nicolis_1971, England2015-sx, Goldbeter2018-et, England2022-xq}. In the operational framework presented here, the cyclic turnover of the molecular assemblies supporting biological function is mapped directly onto the physical definition of a dissipative structure (Fig. 1c). These continuous cycles generate and sustain localized, low-entropy molecular states (e.g., functional proteins) by dissipating driving free energy gradients and exporting entropy to the surrounding environment.

The defining distinction between non-living far-from-equilibrium systems (such as atmospheric vortices) and living systems lies in the self-perpetuating nature of the biological process. In non-living systems, the dissipative structure emerges and vanishes passively, dictated entirely by fluctuating external boundary conditions. In contrast, living systems actively modulate their internal states to preserve the non-equilibrium thermodynamic gradients required for their continued structural persistence---a capacity grounded in their autopoietic organization \citep{maturana2012autopoiesis}. Accordingly, at the systemic level, the fundamental function of the biological network is the continuous, active regeneration of the network itself \citep{varela2025principles}.

Within this autopoietic organization, the metabolic network consists of an intricate cybernetic \citep{wiener1961cybernetics} array of functional molecular species ($M_1, M_2, \dots, M_n$) that orchestrates this energy- and information-driven self-maintenance. However, rather than tracing this full topological complexity, isolating the lifecycle of a single representative functional macromolecule $M$ within its dissipative space reveals the core kinetic trajectory (Fig. 1c). In this reduced organizational representation, the radial axis captures the continuous, vortical metabolic flux driven against the entropy gradient ($\nabla S$). The species ready to execute biological function ($M$) sits at the apex of this turnover loop---the state of maximum local structural order---and its functional execution ($M \rightarrow M^*$) propels its progression down its catabolic slope.

Crucially, it is the conditional availability of these metastable species at the apex---sustained exclusively by proper anabolic regulation---that creates the physical substrate required for cybernetic interactions (Fig. 1d). Indeed, classical static representations of signaling pathways merely reflect projections of what occurs on a plane tangential to this continuous molecular turnover.

The functional execution $M \rightarrow M^*$ (Fig. 1c) continuously depletes the very physical substrate upon which cybernetic signaling relies, generating a standing demand for structural restoration. Physiology is thus forced to operate across a stark kinetic and thermodynamic asymmetry: a rapid, downhill consumption offset by a logistically complex, kinetically slow, and energy-intensive anabolic reconstruction---a fundamental constraint that dictates the limits of biological function.

\section{Neural computation consumes its substrate}

Modern digital computing is celebrated for its accuracy, reliability, and reproducibility. These properties arise from two foundational features: signal stability and the physical inertness---or structural invariance---of the hardware substrate, a stark contrast to the far-from-steady-state dynamics of living neural tissue. In digital systems, information is encoded in discrete states---typically binary voltage levels---that are reliably propagated through processing layers without degradation. The physical media underlying this process, such as silicon and copper, are highly stable; they do not wear down appreciably during computation, and their properties remain essentially static even after years of continuous operation. This material robustness enables complete decoupling of information processing from physical degradation: a numerical simulation can be paused, its state saved, the system powered down, and execution resumed days later as if no time had passed. In essence, digital computing operates in an idealized regime where information is architecturally abstracted from matter.

Before digital dominance, analog computing provided the primary paradigm for solving mathematical problems. Information was represented not by discrete symbols, but by continuous physical variables---such as voltages, currents, or mechanical displacements. While analog systems could model differential equations in real time, they were inherently vulnerable to signal drift, thermal noise, and material degradation. Subtle changes in component properties---such as temperature-dependent shifts in resistance---accumulated over time, distorting computational outputs and demanding frequent calibration. Analog computing thus revealed a fundamental vulnerability: whenever computation depends directly on the physical state of its components, material stability becomes the rate-limiting factor for operational accuracy.

Biological computation shares with analog systems a reliance on continuous, graded signals---such as temporal spike patterns, neurotransmitter gradients, and postsynaptic potentials \citep{Gebicke-Haerter2023-ig}. Yet it introduces a far more radical volatility: its substrate is neither inert nor fixed, but composed of non-equilibrium dissipative structures undergoing continuous physical turnover. In this regime, the distinction between signal and hardware dissolves, as both are embedded within the same stream of structural transformation \citep{Laydevant2024-do}. Unlike any engineered system, biological computation does not merely run upon a physical structure---it explicitly consumes its structural substrate in the very act of execution.

\section{The Self-Calibration of an Autopoietic Analog Computer}
\label{sec:Self-Calibration}

\subsection{Computational Drift in a Dissipative Substrate} 
\label{sec:synaptic-turnover}

The inherent thermodynamic asymmetry of biological function (Fig. 1c) finds a clear illustration in synaptic transmission (Fig. 2a). At the presynaptic terminal, the sub-millisecond exocytosis of a neurotransmitter vesicle represents the rapid collapse of an elaborate, energetically costly preparatory cycle \citep{Sudhof2013-ia}. Vesicle biogenesis, cytoskeletal transport, and high-gradient neurotransmitter packing construct a low-entropy, metastable physical structure continuously reformed and primed at the active zone \citep{Bolz2025-rs}. Upon action potential arrival, this structural and energetic investment is abruptly discharged, displacing the substrate toward a high-entropy, functionally degraded state (Fig. 2a, right panel). Here, the tension between rapid substrate consumption and slow anabolic regeneration manifests as \emph{synaptic fatigue}---alternatively termed \emph{short-term synaptic depression} \citep{Pozzo-Miller1999-mn, Zucker2002-fe, Spano2019-tr}. This framework adopts the classic definition of synaptic fatigue: the reduction in synaptic strength driven by the progressive depletion of functional neurotransmitter vesicles under sustained or high-frequency stimulation \citep{Zucker1989-dz, Dobrunz1997-dw, Pozzo-Miller1999-mn}. Rather than an adaptive form of neural plasticity---a term that often carries teleological implications---fatigue represents a direct physical mismatch between driving input frequencies and the rate-limited kinetics of vesicle replenishment.

A parallel thermodynamic asymmetry governs postsynaptic machinery. Instead of acting as static signal transducers, postsynaptic receptors undergo continuous, non-equilibrium assembly, activation, internalization, and degradation. Sustained receptor activation drives progressive desensitization and endocytic degradation, systematically reducing local synaptic efficacy \citep{Zucker2002-fe, Sheng2002-cp}.

Rather than a binary failure mode, acute synaptic fatigue marks the extreme threshold of a continuous, graded physical deterioration across the synaptic secretory machinery. At the presynaptic terminal, evoked neurotransmission is governed by the vesicular release probability ($P_r$) \citep{Branco2009-nr, Korber2016-ps}---a continuous state variable modulated by the operational history of the active zone, tracing a functional trajectory from optimal release efficiency to presynaptic fatigue. Crucially, $P_r$ has recently been proposed as a universal, cross-species metric tracking sleep need, where its progressive decline during prolonged wakefulness reflects this presynaptic fatigue, gradually impairing neurotransmission efficiency and temporal precision \citep{Wu2025PresynapticReleaseProbability}.

The high computational throughput of the nervous system relies on fast intraneuronal electrical conduction via action potential propagation. Interneuronal communication, however, depends on vesicular secretion---a mechanism that is comparatively slow owing to its sequential, multi-stage lifecycle (requiring neurotransmitter loading, vesicle trafficking, docking, priming, and endocytic retrieval), while remaining energetically costly and logistically demanding \citep{Gupta2025-ua}. Managing the logistical bottleneck and metabolic constraints of neurotransmitter release requires multi-tiered structural buffers---ranging from immediately primed Readily Releasable Pools (RRP) at the active zone to reserve pools organized within biomolecular condensates \citep{Brodin2026-qq}. While local endocytic recycling temporarily prolongs vesicle availability, it operates on an aging molecular substrate continuously accumulating oxidative damage \citep{Beckhauser2016-mc}. Local recycling thus acts as a temporary stopgap rather than a permanent solution: because it cannot halt cumulative structural entropy, sustained computational loads steadily degrade the underlying hardware, creating a structural debt that only \emph{de novo} anabolic turnover can fundamentally reset.

These biophysical constraints are further exacerbated by the extreme spatial and stoichiometric limits of neuronal architecture, which impose complex logistical demands on molecular turnover. Synapses are sub-femtoliter compartments \citep{Harris2012} containing remarkably low copy numbers of essential functional proteins \citep{Sheng2011}, often localized hundreds of micrometers away from the soma. To sustain execution despite structural and logistical bottlenecks, neurons rely on specialized, temporally segregated mechanisms for synaptic machinery turnover. For example, rodent models show that circadian-regulated waves of mRNAs are transported along axons and dendrites to synaptic terminals in two distinct phases: a dusk wave enriched for synaptic transmission components to prepare for active wakefulness and a dawn wave enriched for translation and metabolic machinery to support substrate reconstruction during sleep \citep{Noya2019, Bruning2019-vp}.

\subsection{Sleep: A Systemic Recalibration State}
\label{sec:systemic-recalibration}

Thermodynamic, kinetic, and spatial constraints expose neural computation as an inherently dissipative enterprise subject to continuous parametric drift. Under these constraints, wakefulness cannot maintain a continuous non-equilibrium steady state---a state fundamentally precluded by the dynamic, critical phase transitions that govern self-organizing physiological systems \citep{Bettinger2023-wn}---rendering uninterrupted computation physically unachievable, and thus revealing neural computation as a fundamentally self-limiting process. This intrinsic limitation is further exemplified by high-entropy boundary conditions such as glutamate excitotoxicity \citep{Wu2025-nr} and interstitial waste accumulation \citep{Hauglund2026-rk}, both of which demonstrate how an entropy-accumulating, non-stationary regime degrades systemic functional capacity. From this perspective, sleep emerges as an evolutionarily conserved physiological state acting as the necessary thermodynamic complement to active, stimulus-driven neural computation---an anabolic phase organized specifically to counterbalance cumulative entropic drift and restore substrate integrity.

The massively parallel and spatially distributed nature of neural computation, driven by unpredictable, multimodal information streams, imposes a highly heterogeneous pattern of structural wear across individual network nodes (Fig.~2b). Crucially, because neural circuits represent a paradigmatic architecture of \emph{tightly coupled analog computation}, the operational fidelity of any single node strictly depends on the coordinated calibration of the broader network manifold. Consequently, once a critical threshold of structurally degraded units introduces local signal distortion, the computational integrity of the entire collective degrades. For this reason, sleep cannot be deployed as a piecemeal, localized maintenance routine; it must be implemented as a coordinated, systemic physiological state.

Crucially, the systemic shutdown required for structural restoration demands a dedicated control interface capable of enforcing a global operational state transition. In mammals, this role is fulfilled by well-characterized subcortical sleep-control circuits---primarily the mutually inhibitory flip-flop switch spanning the ventrolateral preoptic nucleus (VLPO) and monoaminergic/orexinergic arousal centers \citep{Saper2010-ih, Sakurai2021-qe}. Rather than acting as the primary driver of sleep necessity, these subcortical networks function as an executive trigger. By integrating physiological surrogates of sustained wakefulness---such as extracellular adenosine accumulation \citep{Reichert2022-vu} and SIK3-mediated cumulative synaptic protein phosphorylation \citep{Funato2016-kc, Wang2018-ar}---they enforce a non-linear state transition that suspends sensory-bound computation, allowing systemic recalibration to occur.

\section{Discussion}

\subsection{Sleep Homeostasis}

Modern definitions of sleep across taxa consistently place homeostatic regulation at the foreground, operationally characterized by a compensatory \emph{sleep rebound} following prolonged wakefulness \citep{Joiner2016-ig, Nath2017-en, Kanaya2020-ly}. To formalize these rebound dynamics, Alexander Borbély introduced the Two-Process Model, operationalizing homeostatic sleep pressure as a continuous variable (\emph{Process S}) that accumulates during wakefulness and decays during sleep \citep{Borbely2022-cs}. By coupling exponential kinetics to dynamic circadian thresholds, Borbély built a remarkably robust framework—in effect, a stiff switching system that captured complex sleep-wake timing and rebound responses with minimal variables. 

Historically, however, the conceptual narrative surrounding this framework has leaned heavily on classical point-attractor homeostasis, describing Process S as a negative-feedback drive seeking a fixed internal set point \citep{Borbely2022-cs}. While empirically transformative, interpreting this architecture through a purely static set-point lens inherits known theoretical constraints \citep[for a comprehensive survey of revised regulatory models in physiology, see][]{Ramsay2014-vl}, such as \emph{rheostasis} \citep{Mrosovsky1990-ef} and \emph{allostasis} \citep{Sterling2012-ai}, which accommodate sliding baselines and predictive adaptation. In this light, part of the theoretical friction in identifying sleep’s core biological substrate may not stem from biological opacity alone, but from interpreting a fundamentally dynamic switching architecture through an overly rigid regulatory metaphor \citep[for a critical evaluation of set-point assumptions in sleep research, see][]{Franken2024-at}.

Following the framework established in Section~\ref{sec:Self-Calibration}, a non-equilibrium dynamical systems perspective reveals that sleep-wake regulation operates without a static point attractor. Rather than a system reluctant to change—stubbornly defending a fixed baseline against external perturbations—neural hardware embodies an architecture designed to consume its own physical substrate during active operation, deferring structural restoration to an offline state. The sleep-wake cycle is thus best conceptualized as a \emph{relaxation oscillator} \citep{Strogatz2019, Skeldon2025-im} driven by multi-scale cybernetic regulation: an integrated control mechanism operating not via static error correction, but through two alternating operational regimes: a dissipative catabolic phase of neural computation that progressively accumulates structural entropic debt, and a complementary anabolic phase organized specifically to settle that debt, calibrate synaptic weights, and restore substrate integrity.

\subsection{The Synapse as the Common Arena of Neural Computation, Plasticity, and Sleep}
\label{sec:synaptic-arena}

The synapse serves as the primary physical arena where computation, plasticity, and metabolic constraints directly intersect \citep{Abbott2004-fs, Lillicrap2020-ie}. This framework foregrounds the ongoing molecular turnover of synaptic machinery, whose far-from-equilibrium volatility drives the drift of fundamental analog computing parameters---a degradation vividly illustrated by synaptic fatigue (Section~\ref{sec:synaptic-turnover}). Crucially, the prevailing paradigm in sleep neuroscience, the Synaptic Homeostasis Hypothesis (SHY) \citep{Tononi2003-dd, Tononi2014-sf, De_Vivo2017-xt}, operates within this exact physical space, though it conceptualizes sleep-dependent restoration almost exclusively through the lens of postsynaptic weight scaling and learning-dependent plasticity.

Significantly, ultrastructural data from the Tononi and Cirelli research program---obtained via serial block-face scanning electron microscopy---demonstrates that enforced wakefulness induces structural changes consistent with SHY, such as enlarged axon--spine interfaces \citep{Spano2019-tr}. However, these same dataset analyses simultaneously expose a progressive depletion in the proportion of readily releasable presynaptic vesicles, an explicit structural signature of presynaptic fatigue. While presynaptic fatigue is a classic physiological phenomenon, it remains notably absent from traditional plasticity-centered models of sleep---a gap explicitly noted in the authors' own discussion \citep[Discussion]{Spano2019-tr}. From a systems perspective, this empirical duality indicates that postsynaptic renormalization as proposed by SHY---typically associated with learning-dependent potentiation in complex brains---represents a specific, evolutionary downstream manifestation of a far more universal recalibration process.

\subsection{Sleep in Animal Evolution}
\label{sec:sleep-evolution}

Neurons achieve fast, low-loss signal propagation along their processes through action potentials, exhibiting quasi-digital reliability. Synaptic signaling, however, relies on vesicular secretion---an evolutionarily ancient mechanism burdened by complex logistics. Established at the foundations of proto-nervous systems ($\sim 600+$ Mya) \citep{Sachkova2025-qr}, the kinetic and thermodynamic constraints imposed by secretory and trafficking machinery within complex neuronal morphologies represent a primary bottleneck in sustained computation, inherited across animal lineages.

Neural structure is inherently dynamic---far from a steady state---yet mainstream frameworks in neural computation routinely subsume these structural shifts under the paradigm of neural plasticity, failing to properly integrate distinct physical phenomena such as biophysical fatigue, molecular degradation, and mandatory protein turnover \citep{Hafner2019-tf, Spano2019-tr, Cirelli2019-io}. As demonstrated by the circadian segregation of synaptic versus translational machinery \citep{Noya2019}, evolution rejected continuous turnover---close to a steady state---in favor of a discontinuous operational schedule. Under the current framework, wakeful computation is engineered to tolerate progressive structural degradation until a system-wide state transition---sleep---enables wholesale anabolic turnover. Crucially, because non-equilibrium biological structures are inherently path-dependent, this restorative turnover cannot execute a simple reset to a prior structural template. While sleep is widely recognized as a critical window for memory consolidation and adaptive function \citep{Diekelmann2010-kd, Tononi2014-sf}, neural plasticity is here subsumed under the thermodynamic imperative of structural turnover, emerging directly as the mechanical byproduct of this discontinuous renewal: each anabolic cycle reconstructs the computational substrate under physical constraints left by prior activity, unifying structural maintenance and adaptive remodeling into a single biophysical event.

Answering Rechtschaffen's influential paradox \citep[p.~88]{Rechtschaffen1971}, sleep is far from an evolutionary mistake---it represents an unavoidable biophysical trade-off, an anabolic window where mandatory molecular turnover and the emergence of neural plasticity inherently collapse into the same physiological process. Within this framework, sleep and sleep-like behaviors are not contingent adaptations tied to specific cognitive functions but systems-level manifestations of conserved biophysical constraints, providing a principled explanation for their ubiquity across the animal kingdom.

\subsection{Empirical Avenues and Testable Signatures}
\label{sec:experimental-avenues}

An out-of-equilibrium thermodynamic perspective predicts an explicit coupling across biological scales: nanoscale structural cycles of disorganization and restoration---driven by the superposition of functional execution and physical turnover---must directly map onto macroscale computational performance, manifesting as progressive operational degradation \citep{Vyazovskiy2011-vc} and sleep-dependent recalibration \citep{Kharas2024-qs}. Although the entropic trace left by active neural computation is delocalized and distributed from subcellular molecular architectures to circuit-level networks, reinterpreting existing experimental paradigms offers immediate entry points to test these predictions---simply by shifting the analytical lens from functionalist scaling to non-equilibrium thermodynamics.

Crucially, this framework predicts that local entropic debt accumulates not as a passive function of time spent awake, but as a direct integral of the dissipative work performed by active circuits. This aligns with empirical observations that sleep pressure accelerates significantly during theta-dominated wakefulness relative to quiet wake \citep{Vassalli2017-bg, Franken2024-at}. Taking this paradigm as a foundation, a key experimental trajectory is to extend this approach by decoupling raw waking duration from cumulative computational load---using EEG features that uniquely mark and scale with active wakefulness, such as the ultrafast component of EEG $\Psi$-patterns \citep{diaz2026recovering}, as a proxy for neural workload.

At the subcellular level, this physical degradation manifests directly when contrasting spontaneous sleep against forced sleep deprivation through a physical lens \citep[e.g.,][]{Spano2019-tr}. Rather than an abstract ``sleep pressure,'' sleep deprivation reflects the progressive driving of sub-synaptic architecture toward the physical limits of its functional stability. Observed ultrastructural alterations under electron microscopy---such as shifts in active zone architecture and synaptic vesicle distribution---mark the measurable trajectory of nanoscale function-structure decoupling. While fully capturing these non-equilibrium kinetics remains an empirical challenge, promising avenues like ``flash-and-freeze'' optogenetic EM \citep{Watanabe2016-tz, Borges-Merjane2020-jr, Imig2020-ck} and emerging cryo-electron tomography of synaptic molecular architecture \citep{Kravcenko2024-dw} point toward a future where nanoscale synaptic structural dynamics can be mapped directly against physiological sleep demand. Indeed, recent work associating sleep need with $P_r$ provides crucial empirical support for this trajectory \citep{Wu2025PresynapticReleaseProbability}, demonstrating that the drop in $P_r$ following prolonged wakefulness correlates directly with ultrastructural active zone disorganization, including altered vesicle density and spatial uncoupling between VGCCs and release sites.

Future experimental efforts toward resolving the enduring mystery of sleep require a fundamental theoretical shift: (i) replacing legacy set-point homeostasis with cybernetic principles aligned with the thermodynamic constraints driving the dissipative dynamics of biological substrates; (ii) recognizing that complex biological networks cannot maintain steady-state regimes, forcing neural computation to rely on a degrading substrate and drift across floating parameters; and (iii) acknowledging that because this parameter drift occurs across tightly coupled nodes, recalibration cannot be executed locally within an active signaling network, but demands a systemic functional suspension.

\clearpage
\section{Figures}

\begin{figure}[htbp]
  \centering
  \includegraphics[width = \textwidth]{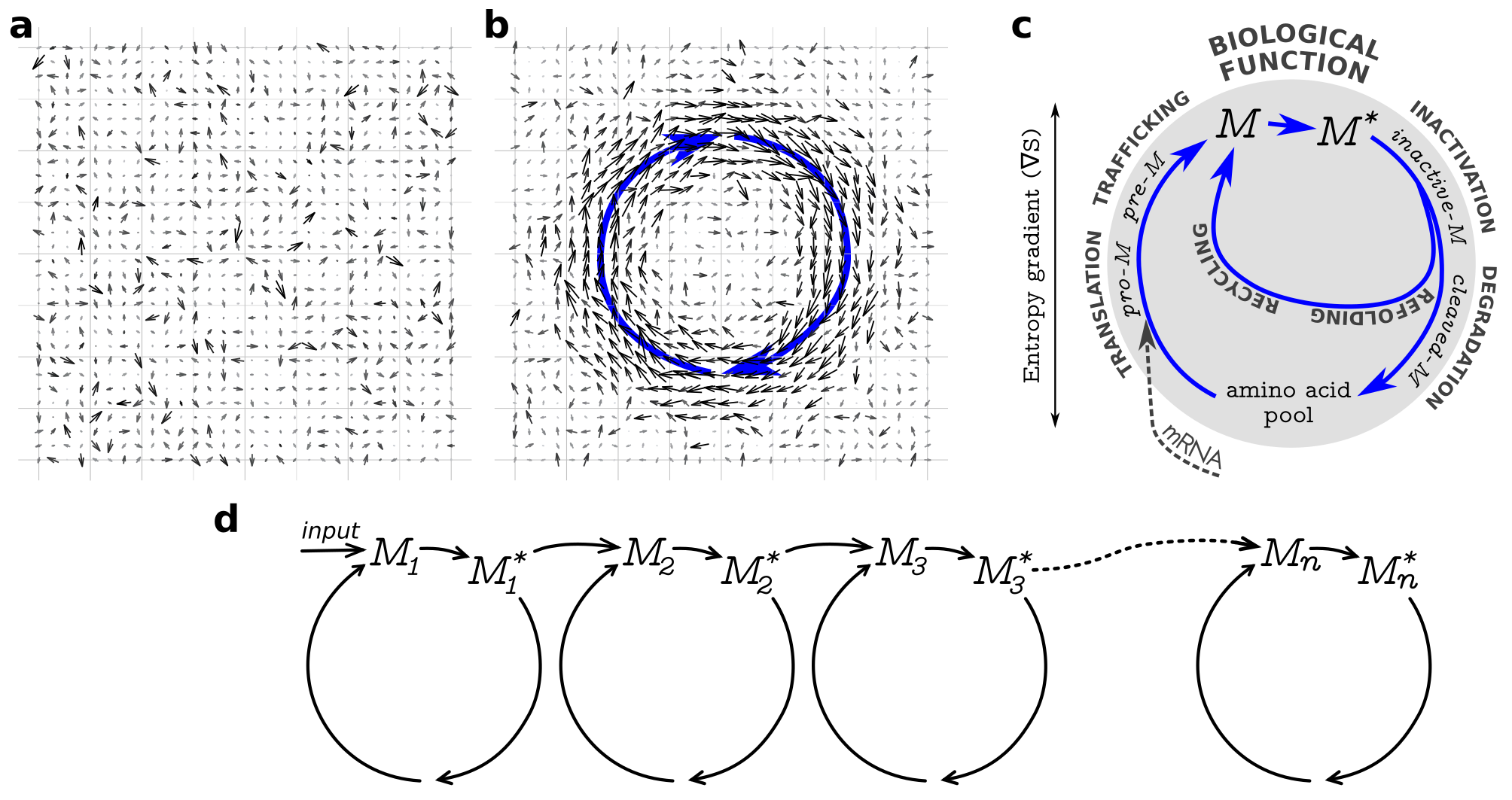}
  \caption{\footnotesize
      \textbf{Systems far from thermodynamic equilibrium.} 
\textbf{a.} At thermodynamic equilibrium, particles move randomly with velocities governed by the Maxwell--Boltzmann distribution. The system exhibits no preferred macroscopic flows and lacks the capacity to perform work. 
\textbf{b.} Sustained energy gradients can break spatial symmetries and induce self-organization (i.e., dissipative structures). A phase-space vector field illustrates a macroscopic vortex-like trajectory (blue arrows), representing common abiotic dissipative structures found in nature, from Rayleigh--Bénard convection cells to tropical cyclones. 
\textbf{c.} Life represents the ultimate manifestation of dissipative structures, where the dynamic, non-equilibrium flow becomes self-contained and self-producing (i.e., autopoietic). The turnover of a single functional macromolecule $M$ is depicted. These metastable species are continuously generated through anabolic processes to execute their biological function ($M \rightarrow M^*$) and are ultimately degraded through catabolic pathways. This microscopic loop is actively driven by a metabolic energy gradient and directed by genetic information ($\text{mRNA}$), forcing the substrate against a steep entropy gradient ($\nabla S$) to sustain local order while the system continuously exports entropy to the environment.
\textbf{d.} Dissipative grounding of sequential signaling. A linear cascade ($M_1 \rightarrow M_2 \rightarrow \dots \rightarrow M_n$) serves as a minimal topological proxy for complex cybernetic networks, demonstrating that inter-molecular signaling depends on the conditional availability of functional species ($M_i$) at the apex of their turnover loop. Upon interacting with upstream components, $M_i$ transitions to an active, executed state ($M_i^*$), propagating the signal tangential to the loop while inexorably driving the substrate down its catabolic slope. Information processing is thus directly coupled to local structural consumption, and hence requires the recycling or resynthesis of participating molecules.}
    \label{fig:biolFunc}
\end{figure}

\clearpage
\begin{figure}[htbp]
  \centering
  \includegraphics[width = \textwidth]{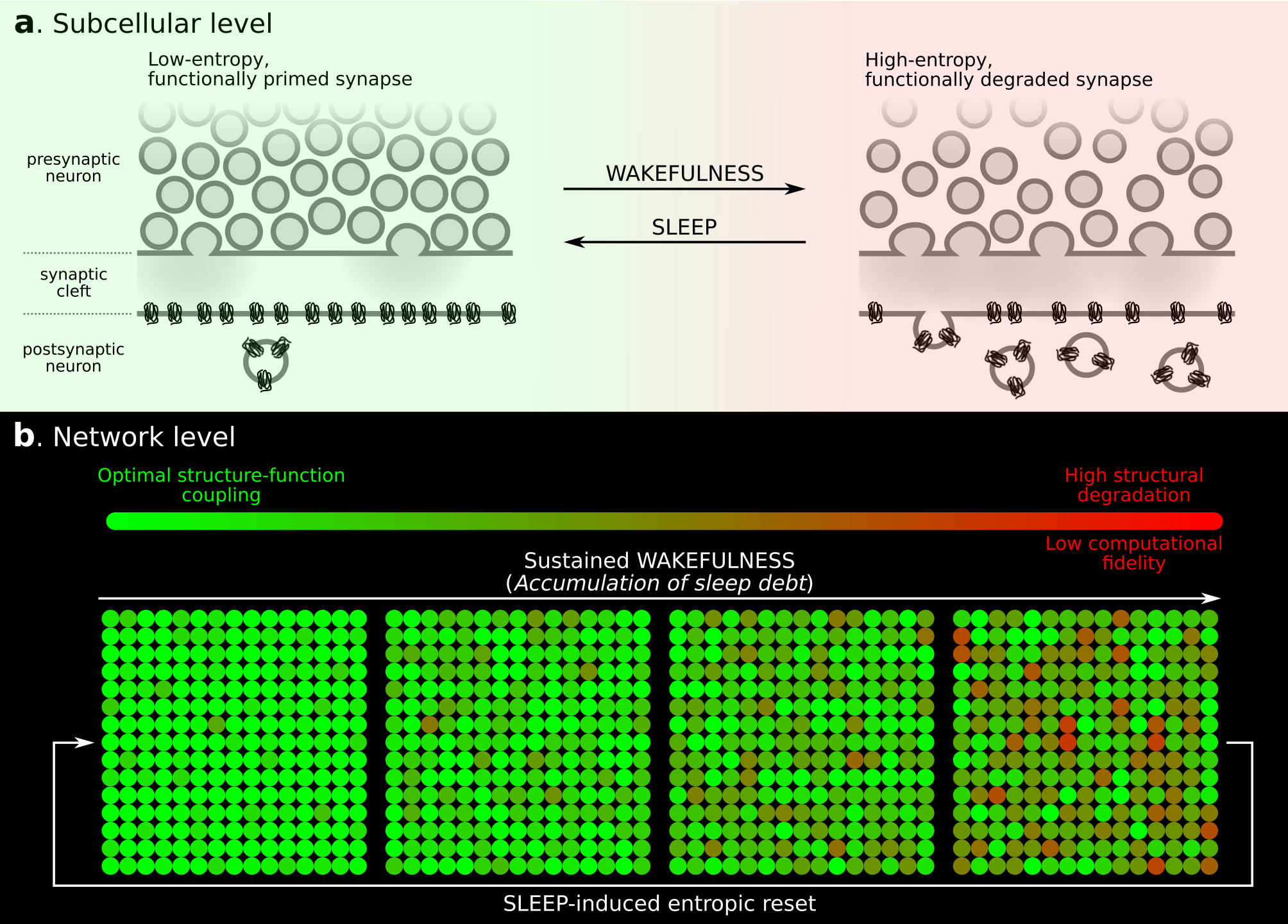}
  \caption{\footnotesize
\textbf{Figure 2. Local structural entropy and network-level computational degradation across wake-sleep cycles.} 
\textbf{a}, \textbf{Subcellular level.} Synaptic fatigue illustrates the critical dependence of functional efficacy on local structural order. During wakefulness, sustained high-demand neural activity depletes pools of low-entropy, functionally primed species (e.g., organized presynaptic vesicle pools, postsynaptic receptor arrays), leading to a high-entropy, functionally degraded state. Sleep restores these low-entropy configurations through anabolic recalibration. 
\textbf{b}, \textbf{Network level.} Abstract representation of a neural network ($15 \times 12$ unit grid) displaying local structural entropy gradients. As wakefulness is sustained (accumulating sleep debt), structural degradation accumulates non-homogeneously due to unpredictable, task-dependent environmental demands. Because of the recurrent nature of neural information processing, signal distortion introduced by even a small fraction of degraded units propagates through the network, ultimately driving a collapse in system-wide computational fidelity. Sleep drives an entropic reset, restoring optimal structure-function coupling across the network.}
    \label{fig:synapse}
\end{figure}

\clearpage
\section*{Funding}

This work was supported by Research Excellence through the International Collaboration and Network Construction (RECONNECT) Grant Number JP22K21351, AMED Moonshot Research and Development Program Grant Number JP21zf0127005, and the World Premier International Research Center Initiative (WPI), MEXT, Japan.
\bibliography{refList}

\end{document}